\documentclass[12pt]{article}
\usepackage{pgfplots}
\usetikzlibrary{decorations.pathmorphing}

\tikzset{snake it/.style={decorate, decoration=snake}}
\pgfplotsset{compat=1.10}
\usepgfplotslibrary{fillbetween}
\usetikzlibrary{patterns}
\usepackage{draft} 

\usepackage{hyperref}
\usepackage{graphicx,color,subfigure}
\usepackage{cite}
\usepackage{mciteplus}
\usepackage{skak}
\usepackage{xcolor}
\usepackage{empheq}
\usepackage{tikz}
\DeclareFontFamily{OT1}{pzc}{}
\DeclareFontShape{OT1}{pzc}{m}{it}{<-> s * [1.10] pzcmi7t}{}
\DeclareMathAlphabet{\mathpzc}{OT1}{pzc}{m}{it}

\def\be#1\ee{\begin{align}#1\end{align}}



\begin{document}

\unitlength = .8mm

\begin{titlepage}

\begin{center}

\hfill \\
\hfill \\
\vskip 1cm

\title{\Huge On Cosmological Singularities in String Theory}

\author{Jinwei Chu and David Kutasov}
\address{
Leinweber Institute for Theoretical Physics,\\  Enrico Fermi Institute, and Department of Physics\\ University of Chicago, Chicago IL 60637
}
\vskip 1cm

\email{jinweichu@uchicago.edu, dkutasov@uchicago.edu}

\end{center}

\abstract{
We study the time evolution of a $3+1$ dimensional spacetime, where space is a large three-sphere, due to small perturbations of the background fields. We focus on two classes of deformations. One corresponds on the worldsheet to time-dependent non-abelian Thirring deformations. The other to perturbations of the radius of the three-sphere. In the former case, we find that small deformations generically lead to big-bang and big-crunch singularities, near which the spacetime becomes highly anisotropic. We argue that string theory likely resolves these singularities. In the latter case, general solutions have the property that the radius of the three-sphere goes to infinity at a finite time, but there are no solutions in which it collapses to zero. We also discuss the interplay of these spacetime properties with the corresponding worldsheet RG flows.}

\vfill

\end{titlepage}

\eject

\begingroup
\hypersetup{linkcolor=black}
\tableofcontents
\endgroup

\vfill
\eject

\section{Introduction}\label{secint}
An interesting open question in string theory is under what conditions the theory leads to a big bang and/or big crunch cosmology, such as the one we appear to live in. There has been some work on this in the past
(e.g. \cite{Nappi:1992kv,Giveon:1992kb,Khoury:2001bz,Elitzur:2002rt,Craps:2002ii,Horowitz:2002mw, Berkooz:2002je, Cornalba:2003kd,Hertog:2004rz,Freedman:2005wx,Elitzur:2005kz,Craps:2007ch,Swanson:2008dt,Barbon:2010gn,Maldacena:2010un,Balthazar:2023oln,Das:2025fkj,Papadopoulos:2025dbz} and references therein), but there is no satisfactory general understanding of this issue. In order to gain such an understanding, it is useful to consider specific models, and try to generalize from them. The main purpose of this paper is to take a step in this direction. 

Our starting point is string theory on the spacetime 
\ie
\label{rts3}
\mathbb{R}_t\times \mathbb{S}^3_k\times {\cal M}_k~.
\fe
Here $\mathbb{R}_t$ is the real line corresponding to the time direction, $\mathbb{S}^3_k$ is a three-sphere of radius $R_k=\sqrt kl_s$, and ${\cal M}_k$ a background that we will for now assume to be described by a unitary compact worldsheet CFT. We will discuss the background \eqref{rts3} in the large $k$ limit, in which the spatial three-sphere is large and one can talk about $3+1$ dimensional dynamics. We will restrict the analysis to the bosonic string, since the issues we will focus on are unrelated to the presence of the usual tachyon of that theory. It is easy to generalize the discussion to the superstring.  

The three-sphere in \eqref{rts3} is described on the worldsheet by an $SU(2)_k$ WZW model. This model has central charge 
\ie
\label{ck}
c_k=3-\frac{6}{k+2}~.
\fe
The $SO(4)$ isometry of the sphere is realized on the worldsheet as an $SU(2)_L\times SU(2)_R$ current algebra of level $k$. We will denote the left and right-moving worldsheet currents by $J^a(z)$ and $\bar J^{\bar a}(\bar z)$, respectively. 

The CFT ${\cal M}_k$ in \eqref{rts3} has central charge 
\ie
\label{cM}
c_{\cal M}=22+\frac{6}{k+2}~.
\fe
If it is compact, \eqref{rts3} corresponds to a $0+1$ dimensional background of the bosonic string, and one expects large infrared fluctuations. However, at large $k$ the three-sphere is large, $R_k\gg l_s$, and the background becomes effectively higher dimensional. Thus, one expects to be able to study the theory \eqref{rts3} in a $1/k$ expansion. 

An interesting question is what happens to the CFT ${\cal M}_k$ in the limit $k\to\infty$. If it remains compact\footnote{By compact CFT, we mean a unitary CFT, in which the $SL(2,\mathbb{R})$ invariant vacuum is normalizable, and the spectrum of excitations is discrete.} in this limit, the background \eqref{rts3} describes $3+1$ dimensional physics.
However, we should note that we do not know of any CFT's with these properties, and it is quite possible that they do not exist. This is a version of the scale separation issue that has been discussed in the AdS context (see e.g.~\cite{Coudarchet:2023mfs} for a review). We leave this question to future work. 

A class of backgrounds of the form \eqref{rts3} that certainly exists is one in which ${\cal M}_k$ includes a non-compact factor, such as a Liouville CFT, or the Euclidean $SL(2,\mathbb{R})/U(1)$ coset. E.g. we can take 
\ie
\label{ncM}{\cal M}_k=SL(2,\mathbb{R})_{k+4}/U(1)\times \mathbb{T}^{20}~.
\fe
This CFT is unitary, but it is not compact, due to the semi-infinite cigar factor. Its spectrum includes a continuum above a gap that goes to zero as $k\to\infty$. Thus, at large $k$ the background \eqref{rts3}, \eqref{ncM} becomes $5+1$ dimensional.  

It is also useful to note for future reference that one can take the central charge $c_{\cal M}$ to be larger than the value \eqref{cM}. This leads to a super-critical theory, in which the extra central charge can be compensated by turning on a timelike linear dilaton. 

The background \eqref{rts3} is static. It describes a large (in string units) three-sphere supported by NS $H$-flux, that exists for all time. It is natural to ask whether it has any instabilities in which the three-sphere collapses to small size. This issue was explored in \cite{Bakas:2006bz}, who found that a certain class of time-dependent deformations of \eqref{rts3} does not lead to such instabilities. A typical example of these backgrounds is one where the radius of the three-sphere depends on time. We will revisit this class of deformations, and add to the discussion of \cite{Bakas:2006bz}.

One of the main goals of this paper is to explore a different class of time-dependent deformations of \eqref{rts3}, that describes backgrounds with cosmological singularities. A major theme of the paper is the relation  between the worldsheet RG in the deformed sigma model on $\mathbb{S}^3$, and the corresponding time-dependent solutions in the effective spacetime theory. We will see that what looks in the effective field theory as a spacetime singularity, is likely resolved in string theory. 

The main tool we will employ is an effective action that we found in a related context in~\cite{Chu:2025boe}. In the next section we briefly describe this action, and then use it in our time-dependent context.

\section{The spacetime EFT for current-current deformations}\label{secrev}

In~\cite{Chu:2025boe}, we started from the spacetime background 
\ie
\label{rtsd}
\mathbb{R}^d\times\mathbb{S}^3_k\times \mathcal{N}_k~,
\fe
where $\mathcal{N}_k$ is assumed to be a compact CFT, chosen such that the total worldsheet central charge is 26, similar to the role of $\mathcal{M}_k$ in (\ref{rts3}). The worldsheet Lagrangian was deformed by
\begin{equation}
\label{Lint}
    \mathcal{L}_\text{int}=-\frac{2}{k}\phi_{a\bar b}(x)J^a\bar J^{\bar b}\ .
\end{equation}
From the worldsheet point of view, $\phi_{a\bar b}(x)$ are generalized non-abelian Thirring couplings, that are taken to depend on the coordinates $x\in \mathbb{R}^d$. In spacetime, they correspond to massless fields on $\mathbb{R}^d$, that transform in the adjoint representation of $SU(2)_L\times SU(2)_R$. 

One of the main results of~\cite{Chu:2025boe} was the derivation, following \cite{Kutasov:1989dt} (see also \cite{Cvetic:2000dm,Sagkrioti:2018abh}), of an effective $d$-dimensional Lagrangian for $\phi_{a\bar b}(x)$ in the large $k$ limit. When parameterized by a complex scalar $\chi$ and a real scalar $\phi$ as
\begin{equation}
\label{aaa}
    \phi_{a\bar b}=\begin{pmatrix}
        -\frac{1}{\sqrt{2}}{\rm Re}\;\chi&-\frac{1}{\sqrt{2}}{\rm Im}\;\chi&0\\
        \frac{1}{\sqrt{2}}{\rm Im}\;\chi&-\frac{1}{\sqrt{2}}{\rm Re}\;\chi&0\\
        0&0&\phi
    \end{pmatrix}\ ,
\end{equation}
this Lagrangian takes the form 
\ie
\label{resclagdil}
L_{\rm eff}=\sqrt{g}e^{-2\Phi}\left[-\mathcal{R}-4(\nabla\Phi)^2+L_K+ \frac{2}{k}V(\phi,\chi,\chi^*)\right]\ .
\fe
Here, $\mathcal{R}$ and $\Phi$ are the $d$ dimensional scalar curvature and dilaton, respectively. The kinetic and potential terms are given by
\begin{equation}
\label{bbb}
    L_K=\frac{(2\pi)^2|\nabla \chi|^2}{(1-2\pi^2|\chi|^2)^2}+\frac{(2\pi)^2(\nabla \phi)^2}{(1-4\pi^2\phi^2)^2}\ ,
\end{equation}
and 
\begin{equation}
\label{ccc}
    V(\phi)=\frac{4}{\alpha'}\left(\frac{4\pi^2|\chi|^2}{(1-2\pi\phi)(1-2\pi^2|\chi|^2)^2}-\frac{1}{(1-2\pi^2|\chi|^2)^2}+1\right)\ .
\end{equation}
The Lagrangian \eqref{resclagdil} -- \eqref{ccc} describes dynamics at the distance scale $\sqrt{k\alpha'}$. At large $k$ this scale 
is much larger than the string length $l_s=\sqrt{\alpha'}$, and the higher-derivative corrections to this Lagrangian can in general be neglected.

An interesting one dimensional subspace of the field space labeled by $\phi,\chi$ \eqref{aaa} is  $\chi=-\sqrt2\phi$. In this subspace, the perturbation (\ref{Lint}) is proportional to $\phi \sum_{a=1}^3J^a\bar J^a$, and the resulting worldsheet theory is the non-abelian Thirring model, with coupling that depends on position in $\mathbb{R}^d$. This is a particularly symmetric subspace of \eqref{aaa}, in which the $SU(2)_L\times SU(2)_R$ symmetry of the undeformed background is broken to a diagonal $SU(2)$. In much of the discussion below we will restrict to this subspace. 

The kinetic and potential terms in the Lagrangian (\ref{resclagdil}), given  by \eqref{bbb}, \eqref{ccc}, take in this subspace the form 
\begin{equation}
\label{kinterm}
   K_\phi\equiv  L_K\Big|_{\chi=-\sqrt2\phi}=\frac{3(2\pi)^2(\nabla \phi)^2}{(1-4\pi^2\phi^2)^2}\ ,
\end{equation}
and 
\begin{equation}
\label{Vchiphi}
    V(\phi)=\frac{64}{\alpha'}\frac{\pi^3\phi^3(1-\pi\phi)}{(1+2\pi\phi)(1-2\pi\phi)^3}~,
\end{equation} 
respectively. Changing to a more convenient coordinate on  field space,
\begin{equation}
\label{defphitilde}
     \tilde\phi=\frac{1}{2}\ln \frac{1+2\pi\phi}{1-2\pi\phi}\ ,
\end{equation}
leads to
\begin{equation}
\label{LKVtphi}
    K_{\tilde\phi}=3(\nabla\tilde\phi)^2\ ,\quad V(\tilde\phi)= \frac{1}{4\alpha'}e^{-2\tilde\phi}(e^{2\tilde\phi}-1)^3(e^{2\tilde\phi}+3)\ .
\end{equation}
As $\phi$ varies from $-1/2\pi$ to $1/2\pi$, $\tilde\phi$ ranges from $-\infty$ to $+\infty$. 

\begin{figure}
	\centering
\includegraphics[scale=0.6]{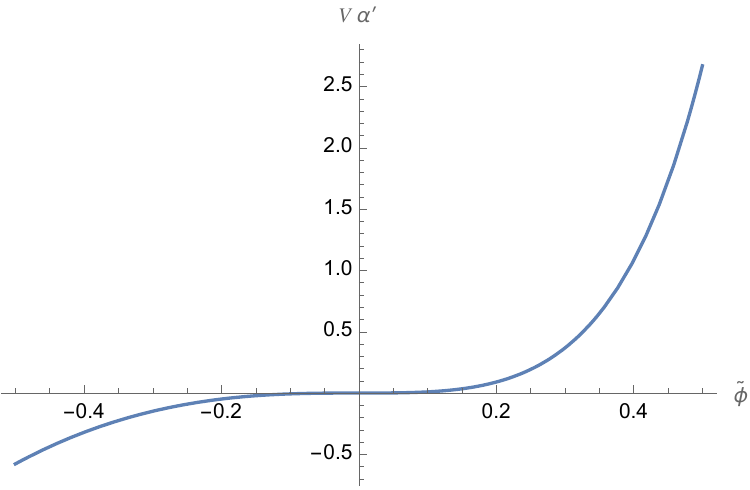}
\caption{\label{Vatphi} The potential $V(\tilde\phi)$ (\ref{LKVtphi}). }
\end{figure}

We plot the potential (\ref{LKVtphi}) as a function of $\tilde\phi$ in figure \ref{Vatphi}. Its structure has a natural worldsheet interpretation. As discussed in 
\cite{Kutasov:1989dt}, it encodes the $\beta$-function of the non-abelian Thirring model via gradient flow, and as such is related to the Zamolodchikov C-function. The Zamolodchikov metric that enters this flow can be read off from the kinetic term \eqref{kinterm}. In the $\tilde\phi$ parametrization, it is constant  (to leading order in $1/k$).

Thus, flowing to the infrared in the worldsheet theory corresponds to moving to smaller $\tilde\phi$ in figure \ref{Vatphi}. Negative $\tilde\phi$ corresponds to the sign of the Thirring coupling where it is marginally relevant. As is well known, in this case the RG flow takes the $SU(2)_k$ WZW CFT to a massive (gapped) one. In particular, the infrared central charge is equal to zero in this case. 

In the $\tilde\phi$ parametrization, the infrared fixed point is pushed to $\tilde\phi=-\infty$. In this limit, the effective potential \eqref{LKVtphi} goes to minus infinity, and the large $k$ approximation that we employed to calculate it breaks down. This is consistent with the fact that in the RG flow from $SU(2)_k$ to a massive theory, the central charge changes by an amount of order one (more precisely, $c_{UV}-c_{IR}=c_k$, \eqref{ck}). In terms of the potential $V$ in \eqref{resclagdil}, this corresponds to 
$V\sim -k$, which is beyond the regime of validity of the calculations in~\cite{Chu:2025boe,Kutasov:1989dt}. From the spacetime point of view, this corresponds to the breakdown of the EFT when the deformed three-sphere becomes stringy, a fact that will be useful below.  

The positive $\tilde\phi$ region in figure \ref{Vatphi} has a qualitatively different physical interpretation. The increase of the potential $V$ with $\tilde\phi$ is consistent with the fact that the non-abelian Thirring perturbation of $SU(2)_k$ WZW is marginally irrelevant in this case. Thus, increasing $\tilde\phi$ corresponds to flowing to the UV on the worldsheet. In general, this introduces an ambiguity having to do with the fact that there is an infinite number of RG trajectories that look in the infrared like a particular irrelevant perturbation of a CFT. 

Geometrically, turning on $\tilde\phi$  corresponds to deforming the round three-sphere of radius $\sqrt{k}l_s$. To leading order in $\tilde\phi$, the deformed three-sphere is described  by the metric
\begin{equation}
\label{ds2}    ds^2=k\alpha'\left(1+2\tilde \phi\right) d\psi^2+k\alpha'\sin^2\psi\left(1+2\tilde\phi\cos2\psi\right)\left( d\theta^2+\sin^2\theta d\varphi^2\right)\ .
\end{equation}
The diagonal $SU(2)$ residual symmetry mentioned above corresponds in the description \eqref{ds2} to the $SO(3)$ symmetry acting on $(\theta,\varphi)$. 

For negative $\tilde\phi$, the perturbation decreases the size of the $\psi$ dimension. It is natural to speculate that in the limit  $\tilde\phi\to-\infty$ the $\psi$ dimension shrinks to zero size. The size of the two-sphere labeled by $(\theta,\varphi)$ may decrease or increase, depending on the value of $\psi$. But even when it decreases, the amount is not enough to maintain a round three-sphere, because $\cos 2\psi<1$. Therefore, roughly speaking, the three-sphere is flattened for $\tilde\phi<0$. Conversely, it is elongated for $\tilde\phi>0$.

\section{Time-dependent solutions}\label{sectphi}

In \cite{Chu:2025boe,Chu:2025kzl} we used the effective action \eqref{resclagdil} -- \eqref{ccc} to study Horowitz-Polchinski solutions, and took the fields $\phi$, $\chi$ to depend on the radial coordinate on $\mathbb{R}^d$. In this paper, we discuss time-dependent solutions. To do that, we replace $\mathbb{R}^d$ in \eqref{rtsd} by $\mathbb{R}^{d-1}\times \mathbb{R}_t$. The background \eqref{rts3} corresponds to $d=1$, and we will mainly focus on this case. We will also restrict to the special case $\chi=-\sqrt2\phi$ discussed above, and study the Lagrangian (\ref{defphitilde}), (\ref{LKVtphi}).

With the metric $g_{tt}$ set to  $-1$ (so that $t$ is the proper time), the Lagrangian (\ref{LKVtphi}) becomes
\ie
\label{LefftphiPhi}
L_{\rm eff}=e^{-2\Phi}\left(-4\dot\Phi^2+3\dot{\tilde\phi}^2- \frac{2}{k}V(\tilde\phi)\right)\ .
\fe
The corresponding e.o.m. are
\ie
\label{Phieom}
4\ddot\Phi=4\dot\Phi^2+3\dot{\tilde\phi}^2- \frac{2}{k}V(\tilde\phi)\ ,
\fe
\ie
\label{tphieom}
\ddot{\tilde\phi}=2\dot{\Phi}\dot{\tilde\phi}-\frac{1}{3k}V'(\tilde\phi)\ .
\fe
Equation \eqref{tphieom} can be thought of as describing the motion of a particle on the $\tilde\phi$ line, in the potential $V(\tilde\phi)/3k$, with the term $2\dot{\Phi}\dot{\tilde\phi}$ acting as (anti-)friction when $\dot\Phi<0(>0)$.

Varying the Lagrangian (\ref{LKVtphi}) w.r.t. $g_{tt}$ gives the Hamiltonian constraint 
\begin{equation}
\label{H0}
    -4\dot\Phi^2+3\dot{\tilde\phi}^2+ \frac{2}{k}V(\tilde\phi)=0\ .
\end{equation}
One can view equations (\ref{tphieom}), (\ref{H0}) as independent, and (\ref{Phieom}) as a consequence of them. To see that, one can take the time derivative of (\ref{H0}) and  replace the $\ddot{\tilde\phi}$ in the resulting equation by (\ref{tphieom}). This yields
\begin{equation}
\label{ddotPhi}
2\ddot\Phi=3\dot{\tilde\phi}^2\ ,
\end{equation}
which combined with (\ref{H0}) gives (\ref{Phieom}). 

Note that while the Lagrangian \eqref{LefftphiPhi} is invariant under time reversal, time-dependent solutions break this symmetry. Indeed, a time reversal invariant solution would have $\dot{\tilde\phi}(0)=\dot\Phi(0)=0$, which implies, via (\ref{H0}), that $V(\tilde\phi(0))=0$. This in turn means that $\tilde\phi(0)=0$, and consequently $\tilde\phi=0$ for all time. 

We next turn to a discussion of solutions of the equations of motion \eqref{tphieom} -- \eqref{ddotPhi}. One can show that all solutions pass through the origin, $\tilde\phi=0$. Therefore, we set $\tilde\phi(0)=0$, and pick a non-zero value for $\dot{\tilde\phi}(0)$. Without loss of generality, we can take $\dot{\tilde\phi}(0)<0$, since solutions with $\dot{\tilde\phi}(0)>0$ can be obtained by taking $t\to -t$. Given this initial data, we can determine $\dot\Phi(0)$ from (\ref{H0}). There are two solutions, $\dot\Phi(0)=\pm \frac{\sqrt3}{2}\dot{\tilde\phi}(0)$. Thus, solutions are labeled by $\dot{\tilde\phi}(0)$ and the sign of $\dot\Phi$.

One general feature of the resulting solutions is that they go to $\tilde\phi\to-\infty$ at a finite time. Indeed, looking back at \eqref{tphieom}, we see that with the above initial conditions, $\tilde\phi(t)$ is negative for $t>0$, and so is $V(\tilde\phi)$ \eqref{LKVtphi}. Equation \eqref{H0} implies that 
\begin{equation}
    \sqrt3 |\dot{\tilde\phi}|\ge \sqrt{-\frac{2}{k}V(\tilde\phi)}\ ,
\end{equation}
or
\begin{equation}
\label{gedt}
    \sqrt{\frac{3k}{-2V(\tilde\phi)}} |d\tilde\phi|\ge |dt|\ .
\end{equation}
Substituting the expression of $V(\tilde\phi)$ (\ref{LKVtphi}) into the l.h.s., one finds that the time it takes for $\tilde\phi$ to reach (negative) infinity is finite.\footnote{The above argument also implies that any solution of \eqref{tphieom}, \eqref{H0}, has the property that if it approaches $\tilde\phi=-\infty$, it does so at a finite time.}

\begin{figure}
	\centering
    \subfigure[]{
	\begin{minipage}[t]{0.45\linewidth}
	\centering
	\includegraphics[width=2.8in]{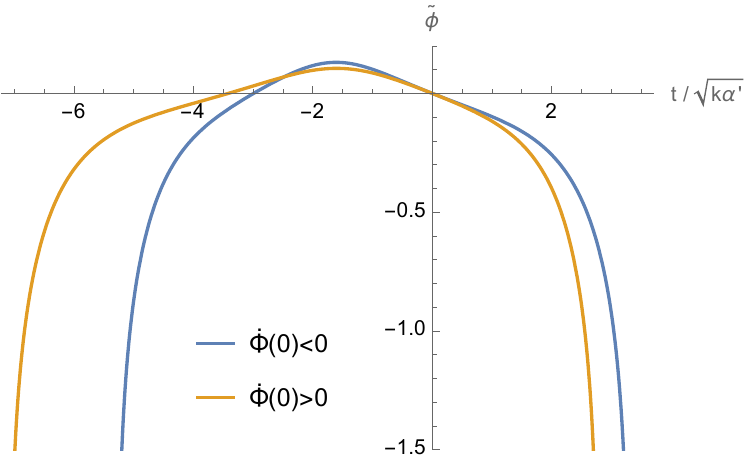}
	\end{minipage}}
	\subfigure[]{
	\begin{minipage}[t]{0.45\linewidth}
	\centering
	\includegraphics[width=2.8in]{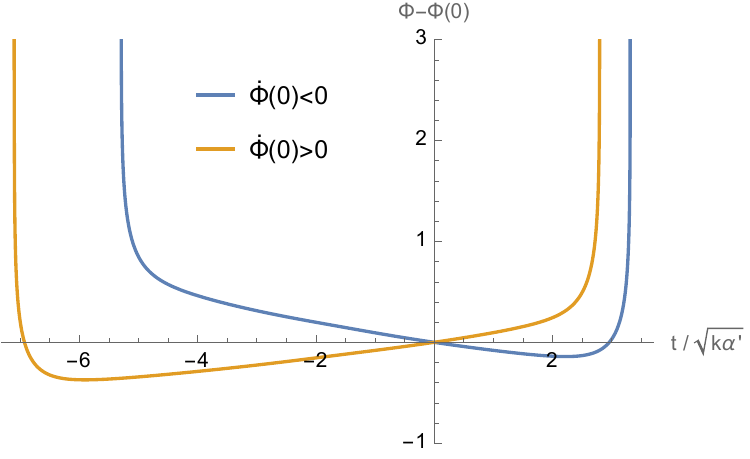}
	\end{minipage}}
	\centering
\caption{\label{tphip0m01}Solutions of (\ref{tphieom}) and (\ref{ddotPhi}) with $\tilde\phi(0)=0$ and $\dot{\tilde\phi}(0)=-0.1/\sqrt{k\alpha'}$. The value of $\dot{\Phi}(0)$ is determined from (\ref{H0}). The singularities occur at $(t_{\text min},t_{\text max})=\sqrt{k\alpha'}(-5.29,3.32)$ for $\dot\Phi(0)<0$, and $\sqrt{k\alpha'}(-7.10,2.80)$ for $\dot\Phi(0)>0$.}
\end{figure}

In figure \ref{tphip0m01} we exhibit an example of the resulting solutions, for the special case $\dot{\tilde\phi}(0)=-0.1/\sqrt{k\alpha'}$. We see that there is no qualitative difference between positive and negative $\dot\Phi(0)$. In both cases the solution describes a spacetime that exists for a finite amount of time, starting at $\tilde\phi=-\infty$ at an initial time, $t_{\text min}$, approaching the vicinity of $\tilde\phi=0$, and then collapsing back to $\tilde\phi=-\infty$ at a final time $t_{\text max}$. Recall that $\tilde\phi=-\infty$ is the EFT representation of the IR fixed point of the worldsheet RG flow, at which the degrees of freedom corresponding to the three-sphere disappear. As mentioned earlier, the vicinity of these singularities are the places where the EFT  breaks down, and one needs to consider the full string theory. 

As we see in figure \ref{tphip0m01}(b), the lower dimensional string coupling $e^\Phi$ diverges at the singularities. Thus, the effective description \eqref{LefftphiPhi} breaks down near the singularities for two reasons. One is the divergence of the potential $V$; the other, the divergence of the string coupling. However, in the full string theory the solution is not necessarily singular there.

\begin{figure}
	\centering
    \subfigure[]{
	\begin{minipage}[t]{0.45\linewidth}
	\centering
	\includegraphics[width=2.8in]{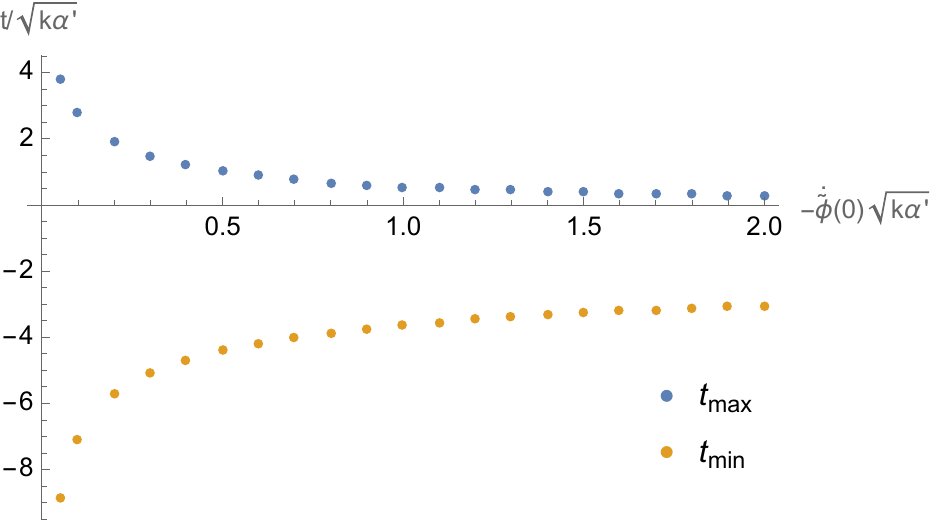}
	\end{minipage}}
	\subfigure[]{
	\begin{minipage}[t]{0.45\linewidth}
	\centering
	\includegraphics[width=2.8in]{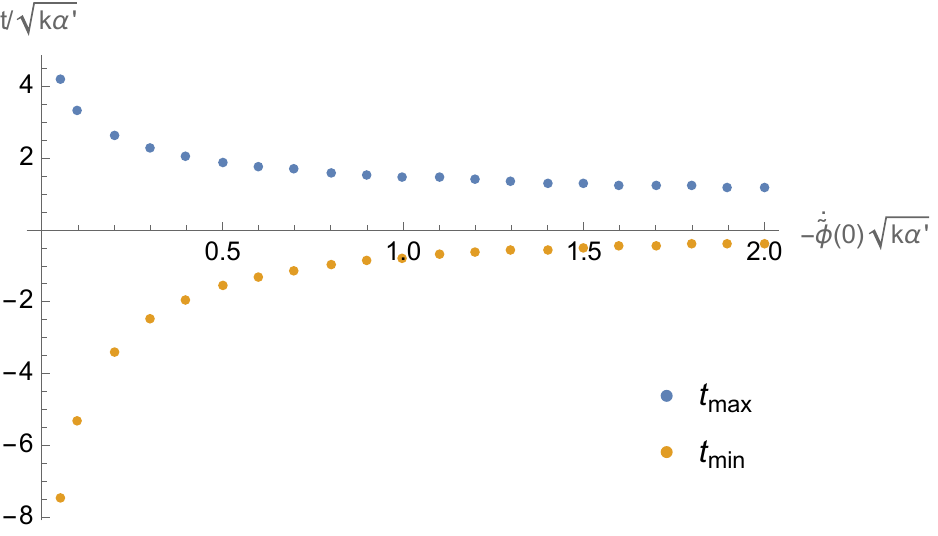}
	\end{minipage}}
	\centering
\caption{\label{tmaxmin}The dependence of the lifetime of the universe, $t_\text{max}-t_\text{min}$, on the initial ``velocity'' $\dot{\tilde\phi}(0)$ for (a) $\dot \Phi(0)>0$, (b) $\dot \Phi(0)<0$.}
\end{figure}

As mentioned above, the solutions of the EFT are labeled by the initial velocity $|\dot{\tilde\phi}(0)|$. In figure \ref{tphip0m01} we chose a particular value for this parameter, but it is interesting to analyze what happens when we vary it. In figure \ref{tmaxmin} we exhibit this dependence. We see that as this parameter becomes smaller, the lifetime of the universe grows, while when it increases, the lifetime approaches a particular finite value. 

One might wonder if increasing $|\dot{\tilde\phi}(0)|$ eventually leads to $\tilde\phi\to\infty$ at $t_\text{min}$. Consider the time reversed solution, namely with $\dot{\tilde\phi}(0)>0$. Presumably, a positive $\dot\Phi$ leads to anti-friction (see (\ref{tphieom})), which would allow $\tilde\phi$ to climb the potential of figure \ref{Vatphi} all the way to infinity if the initial velocity $\tilde\phi(0)$ is large enough. However, we find that this never happens. Indeed, as proven in appendix \ref{exp}, the ability of a field $f$ to climb a potential $W(f)$ which grows exponentially as $f\to +\infty$ depends on the exponent. If the exponent is below a critical value, $2\sqrt3$, the field would be able to climb the infinite potential wall with the aid of anti-friction. Conversely, when the exponent exceeds this critical value, the field cannot escape to infinity despite the presence of anti-friction.\footnote{A similar dependence on a critical exponent for scalar potentials in cosmology has been discussed in previous literature; see, e.g.,~\cite{Halliwell:1986ja,Burd:1988ss,Emparan:2003gg,Wohlfarth:2003kw,Russo:2004ym,Dudas:2010gi}.}
Looking back at (\ref{LKVtphi}), we see that in our case the exponent is equal to six, so $\tilde\phi$ cannot go to infinity at $t_\text{min}$.

To summarize the results of this section, we found that the string background \eqref{rts3} has a family of time-dependent deformations, that can be thought of from the worldsheet point of view as time-dependent non-abelian Thirring deformations of the $SU(2)$ level $k$ WZW model. At large $k$ one can study these deformations using the EFT of~\cite{Chu:2025boe,Kutasov:1989dt}, reviewed in section \ref{secrev}.  As presented in figure \ref{tphip0m01}, the solutions give rise to big bang/big crunch type cosmologies, where the initial and final singularities correspond to backgrounds in which the three-sphere in \eqref{rts3} effectively disappears.

Near the singularities in figure \ref{tphip0m01}, the EFT description of the time-dependent background breaks down, however it is likely that the full string theory description remains non-singular. In the corresponding worldsheet RG, in this region of coupling space the three-sphere degrees of freedom become massive and disappear. The analog of that in the time-dependent solution is an interesting open problem, that will be left for future work.

\section{Isotropic cosmology}\label{secSO4}

The time-dependent backgrounds studied in the previous section have the property that the $SO(4)=SU(2)_L\times SU(2)_R$ isometry of the three-sphere is broken to a subgroup, with the maximal subgroup being the diagonal $SU(2)$. It is natural to ask whether the cosmological singularities we found also occur when the symmetry remains unbroken. In this section we will discuss this issue. 

The only $SO(4)$ preserving deformation is one in which the radius of the three-sphere depends on time. From the worldsheet point of view, it corresponds to adding to the WZW Lagrangian a term proportional to~\cite{Knizhnik:1984nr}  
\begin{equation}
\label{fff}
   \delta R(t)\sum_{a,\bar b=1}^3 J^a\bar J^{\bar b} V_1^{a\bar b}\ ,
\end{equation}
where $V_1$ is the primary field transforming in the spin one representation under the left and right-moving  $SU(2)$'s. From the point of view of the $3+1$ dimensional theory on $\mathbb{R}_t\times \mathbb{S}^3_k$,  \eqref{fff} describes a metric deformation that preserves the full $SO(4)$ symmetry. From the $0+1$ dimensional point of view, it corresponds to a massive excitation, with a mass of order $m_s/\sqrt k$. This has to do with the fact that $V_1$ has scaling dimension $\Delta_1=\bar\Delta_1=2/(k+2)$. Thus, the vertex operator \eqref{fff} is an irrelevant deformation of the sigma model on the three-sphere, in contrast with \eqref{Lint}, which is marginal. 

To study the deformation \eqref{fff} at large $k$, one can use the leading order in $\alpha'$ approximation (see, e.g.,~\cite{Bakas:2006bz}). Denoting the time-dependent radius by 
\begin{equation}
\label{sigdef}
    R(t)=\sqrt{k\alpha'}e^{\sigma(t)}\ ,
\end{equation}
the analogs of (\ref{tphieom}) and (\ref{ddotPhi}) take the form
\begin{equation}
\label{sigeom}
\begin{split}
   \ddot\sigma=&2\dot\Phi\dot\sigma-\frac{1}{6}U'(\sigma)~,\\
   2 \ddot \Phi=&3\dot\sigma^2
    \ .
\end{split}
\end{equation}
Here, $\Phi$ denotes the lower-dimensional dilaton after compactification on the three-sphere as before. It is related to the 3+1-dimensional dilaton, $\Phi_{3+1}$, by
\begin{equation}
\label{dilhi}
\Phi = \Phi_{\text{3+1}} - \frac{3}{2}\sigma~.
\end{equation}

\begin{figure}
	\centering
\includegraphics[scale=0.6]{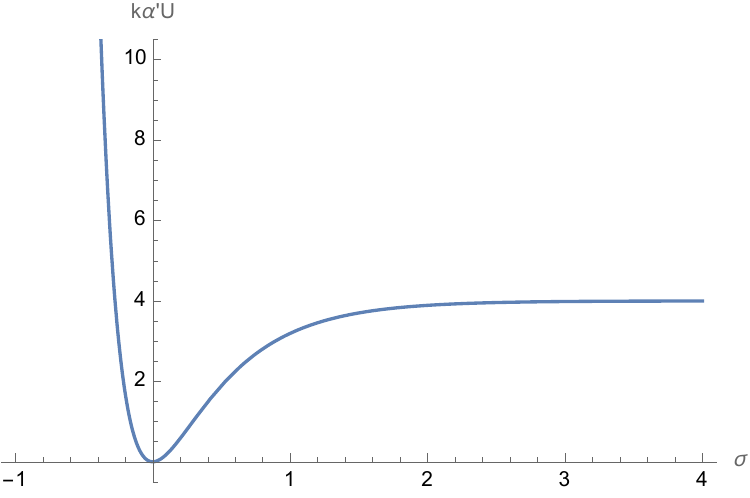}
\caption{\label{Uasig}The potential $U(\sigma)$ (\ref{Usig}). }
\end{figure}

The potential $U(\sigma)$ in (\ref{sigeom}) takes the form
\begin{equation}
\label{Usig}
U(\sigma) = \frac{2}{k\alpha'}\left(-3e^{-2\sigma} + e^{-6\sigma} + 2\right)\ ,
\end{equation}
and is plotted in figure \ref{Uasig}. This potential exhibits a minimum at $\sigma=0$, which is consistent with the fact that the worldsheet perturbation \eqref{fff} is irrelevant for both signs of $\delta R$ (or, equivalently \eqref{sigdef}, $\sigma$). This should be contrasted with the potential of figure \ref{Vatphi}, where the corresponding point, $\tilde\phi=0$, is a saddle point, due to the fact that the non-abelian Thirring perturbation is marginally relevant and irrelevant, respectively,  for the two signs of the coupling. 

As $\sigma\to\infty$, the potential \eqref{Usig} approaches a finite constant. This is consistent with the fact that in this limit the radius of the three-sphere \eqref{sigdef} goes to infinity, and the worldsheet theory approaches a sigma model on $\mathbb{R}^3$. The central charge changes by an amount of order $1/k$ along this RG flow, from \eqref{ck} to $c=3$, in agreement with the fact that the potential $U$ increases by an amount of order $1/k$ between $\sigma=0$ and $\sigma=\infty$.

As $\sigma \to -\infty$, the potential \eqref{Usig} diverges. This limit corresponds to the radius of the three-sphere going to zero, in which the sigma model becomes highly curved and there is no obvious UV fixed point. And, as in our discussion in the previous section, the description above breaks down at $U$ of order one, or $\sigma\sim -\ln k$. 

The Hamiltonian constraint analogous to (\ref{H0}), takes in this case the form
\begin{equation}
\label{H0sig}
    -4\dot\Phi^2+3\dot\sigma^2+U(\sigma)=0\ .
\end{equation}
A consequence of this equation, combined with the fact that $U(\sigma) > 0$ for all $\sigma \neq 0$, is that $\dot\Phi$ cannot vanish, except for the trivial static solution $\sigma(t) = 0$. Thus, for any non-trivial solution, the dilaton field provides either  friction ($\dot\Phi<0$), or anti-friction ($\dot\Phi>0$),  for all time. In the subsequent analysis, we focus on the case $\dot\Phi<0$; solutions with $\dot\Phi>0$ can be obtained from these by the transformation $t\to -t$. 

As before, solutions of \eqref{sigeom}, \eqref{H0sig} pass through the minimum $\sigma=0$ of the potential $U(\sigma)$. This allows us to set $\sigma(0)=0$ without loss of generality. Solutions are then parametrized by the ``initial velocity'' $\dot{\sigma}(0)$. The constraint (\ref{H0sig})  determines the initial dilaton velocity as $\dot\Phi(0)=- \frac{\sqrt3}{2}|\dot{\sigma}(0)|$. 

Before proceeding to a numerical analysis, we first examine the general qualitative behavior of the solutions. In the absence of the friction term in (\ref{sigeom}), at late times the solution $\sigma(t)$ would either oscillate around the origin (for small $\dot\sigma(0)$) or escape to positive infinity (for sufficiently large $\dot\sigma(0)$). When friction is present, the oscillatory solutions become damped. One interesting question is whether $\sigma$ can still escape to $+\infty$.

Assuming that $\sigma\to+\infty$ as $t\to+\infty$, from (\ref{H0sig}) we see that the friction coefficient  $\dot\Phi$ must remain finite in this limit. Consequently, $\dot\sigma$ must decay to zero. Solving (\ref{H0sig}) asymptotically then gives $\dot\Phi\to -1/\sqrt{k\alpha'}$. Substituting this limiting value into the first equation of (\ref{sigeom}) yields the late-time behavior
\begin{equation}
    \sigma\sim A-Be^{-2t/\sqrt{k\alpha'}}\ ,
\end{equation}
where $A$ and $B$ are constants. As $t\to+\infty$, this expression approaches a finite value $A$, contradicting the original assumption of unbounded growth. Therefore, $\sigma$ inevitably oscillates around the minimum $\sigma=0$ with damping. Note that this oscillatory behavior allows us to set $t=0$ such that $\dot\sigma(0)>0$ without loss of generality.

We now analyze the general early-time behavior of solutions. Evolving the system backward in time, a negative $\dot\Phi$ acts as anti-friction. This anti-friction continuously injects energy into the $\sigma$ field, potentially enabling it to overcome the potential barrier and escape to infinity. In fact, if $\sigma\to+\infty$ as $t$ decreases, it must do so in finite time.

To demonstrate this, we first note that approaching $\sigma\to +\infty$ only requires overcoming a finite potential barrier. In this situation, the anti-friction (in the backward time direction) causes the velocity $\dot\sigma$ to grow without bound, eventually dominating the finite potential term in the Hamiltonian constraint (\ref{H0sig}). Consequently (\ref{H0sig}) simplifies to
\begin{equation}
\label{dPhidsig}
    \dot\Phi=\frac{\sqrt3}{2}\dot\sigma \ ,\quad \sigma\to +\infty\ .
\end{equation}
Substituting this relation into (\ref{sigeom}), and neglecting the subleading potential gradient on the r.h.s. yields the asymptotic solution
\begin{equation}
    e^{-\sqrt3\sigma}=Ct+D\ ,\quad \sigma\to +\infty\ ,
\end{equation}
where $C,D<0$ are integration constants.\footnote{The condition $C<0$ follows from the e.o.m.. We additionally impose $D<0$ in order for $\sigma\to+\infty$ to occur in the past ($t<0$).} This solution explicitly shows that $\sigma$ goes to $+\infty$ at the finite time $t=-D/C$.
\begin{figure}
	\centering
    \subfigure[]{
	\begin{minipage}[t]{0.45\linewidth}
	\centering
	\includegraphics[width=2.8in]{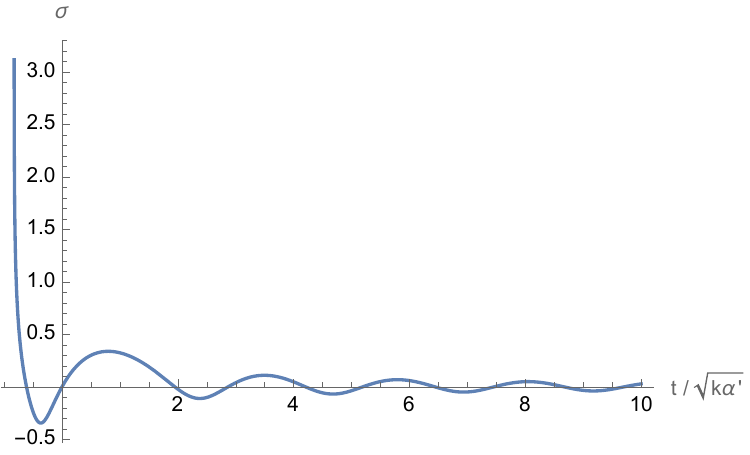}
	\end{minipage}}
	\subfigure[]{
	\begin{minipage}[t]{0.45\linewidth}
	\centering
	\includegraphics[width=2.8in]{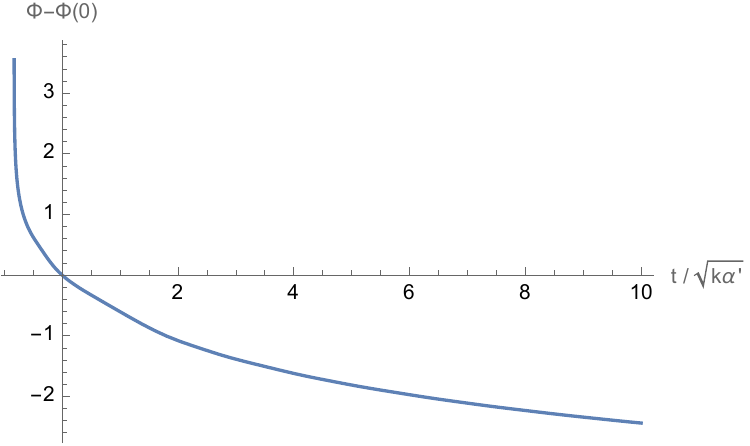}
	\end{minipage}}
	\centering
\caption{\label{sigp01}Solutions of (\ref{sigeom}) with $\sigma(0)=0$ and $\dot{\sigma}(0)=1/\sqrt{k\alpha'}$. The value of $\dot{\Phi}(0)$ is determined from (\ref{H0sig}) using the negative root.}
\end{figure}

In figure \ref{sigp01}, we plot a typical solution, which shows that $\sigma$ originates from $+\infty$ at a finite time and oscillates around the origin with damping. Physically, this describes a three-sphere contracting from an infinite size, undergoing infinitely many oscillations about the critical radius $\sqrt{k\alpha'}$, and eventually settling at this size.

Furthermore, figure \ref{sigp01} shows that at the beginning of time where $\sigma\to +\infty$, the lower dimensional dilaton $\Phi$ goes to $+\infty$, consistent with (\ref{dPhidsig}). According to (\ref{dilhi}), the 3+1-dimensional dilaton goes to infinity as well. In this regime, while the potential $U(\sigma)$ remains valid, the solution breaks down due to the divergence of the string coupling.

In principle, a solution could also originate from $\sigma \to -\infty$, thereby probing an unknown UV worldsheet theory. However, as shown in appendix \ref{exp}, the potential (\ref{Usig}) grows too fast for this to happen.

To summarize, we find two classes of solutions corresponding to time-dependent radius deformations of the three-sphere, distinguished by the sign of $\dot\Phi$. For $\dot\Phi<0$, the three-sphere originates from an infinite size at a finite initial time and undergoes damped oscillations around the radius $\sqrt{k\alpha'}$. 

For $\dot\Phi>0$, obtained from the $\dot\Phi<0$ case via time reversal $t\to -t$, the three-sphere starts at the radius $\sqrt{k\alpha'}$ in the far past. As time progresses, its radius begins to oscillate around this value with increasing amplitude. Eventually, after a final oscillation of sufficiently large amplitude, the sphere expands to infinite size without subsequent recollapse. Interestingly, starting at a time $t=0$ with a finite velocity $\dot\sigma(0)$, this expansion from $\mathbb{S}^3$ to $\mathbb{R}^3$ is completed within a finite time.

Notably, we find no solutions in this setup that exhibit a big-bang or big-crunch singularity, which would correspond to $\sigma\to-\infty$ at early or late times.

\section{Discussion}

In this paper, we studied the time evolution of a class of $3+1$ dimensional spacetimes in string theory. The starting point of the discussion was the  static spacetime $\mathbb{R}_t\times \mathbb{S}^3_k$ \eqref{rts3}, where the spatial three-sphere is described by an $SU(2)$ WZW model with a large level $k\gg1$. We studied the fate of this spacetime under deformations of the size and shape of the three-sphere. 

Our primary focus was on the time-dependent current-current deformations \eqref{Lint}. To analyze them we used the effective spacetime Lagrangian that was derived in~\cite{Chu:2025boe}. We focused on the subspace of this space of deformations in which the $SO(4)$ isometry of the sphere is broken to a diagonal $SU(2)$. From the worldsheet point of view, it corresponds to a non-abelian Thirring model with a time-dependent coupling. 

We found that for small deformations of the static background \eqref{rts3}, the solution has both 
early and late time singularities (see figures \ref{tphip0m01}, \ref{tmaxmin}). Near these singularities, the space becomes small and rapidly varying, and the EFT description breaks down. There are reasons to believe that in the full string theory the singularities are resolved, since the vicinity of the singularities corresponds from the worldsheet point of view to the region in coupling space, where the non-abelian Thirring theory approaches its IR fixed point. That fixed point is well understood, and it is natural to expect that the singularities we found are due to the breakdown of the EFT, and not of the full string theory.  We leave a detailed study of this problem to future work. 

We also discussed the relation of the problem studied in this paper to one where the full $SO(4)$ isometry of the three-sphere remains unbroken. In that case, the three-sphere remains round with a radius that depends on time. This problem 
was previously studied in~\cite{Bakas:2006bz}, and we slightly extended their analysis. We found that  solutions develop a singularity either in the past or in the future—but not both—depending on the sign of the time derivative of the dilaton. These singularities correspond to the three-sphere expanding to infinite size at a finite time. There are no solutions in which the radius of the three-sphere goes to zero. 

The resulting picture is somewhat counterintuitive. One might think that if the three-sphere in \eqref{rts3} is unstable to collapse, the most unstable mode will be the spherically symmetric one. But, in fact, we found that the unstable modes break $SO(4)$ to at most  $SU(2)$. The solution near the big bang/big crunch singularity is highly anisotropic, while near the maximum of the size of the universe it can be isotropic to a very good approximation. It would be interesting if a similar mechanism was in play in our universe.  

It is interesting to ask why the instability structure has the properties summarized above. We do not have a full answer to this question, but the following elements probably play a role. The current-current deformation \eqref{Lint} is marginal (on the worldsheet), and in a particular subspace of coupling space is marginally relevant. On the other hand, the $SO(4)$ invariant deformation \eqref{fff} is  irrelevant. Thus, flowing towards a big bang/big crunch singularity corresponds to flowing down the worldsheet RG in the former case, and up the RG in the latter.

There are several interesting directions for future work. One natural question concerns the observables in the time-dependent backgrounds studied in this paper. In string theory on the static background \eqref{rts3}, one can define standard asymptotic observables associated with $t\to\pm\infty$, at least at large $k$. However, in the big bang/big crunch spacetimes like that of figure \ref{tphip0m01}, these asymptotic regions are absent. It would be interesting to understand what replaces the usual observables in these spacetime. Recent work in this direction in a different context appeared in~\cite{Balthazar:2023oln,Das:2025fkj}. It would be interesting to generalize the analysis of these papers to the present context.

In this paper we treated the dynamics of the fields $\tilde\phi$ and $\sigma$ separately. It could be that allowing both of them to be turned on leads to new phenomena. E.g., it could be that the positive mass squared of $\sigma$ receives negative corrections due to the coupling to $\tilde\phi$, which make it tachyonic. Similarly, the vanishing mass squared of $\tilde\phi$ may receive negative contributions due to the coupling to $\sigma$. In fact, one can think of \eqref{Lint} and \eqref{fff} as the leading terms in the expansion of metric perturbations on a large three-sphere in spherical harmonics. It would be interesting to generalize our description to higher spherical harmonics. This
important problem is also left to future work.

Another set of open questions is associated with the compact CFT 
$\mathcal{M}_k$ in  (\ref{rts3}). As we mentioned in section \ref{secint}, scale separated CFT's with the requisite properties may not exist. In that case, one can consider different options, each of which has its own properties. One option is to consider a non-compact space, such as (\ref{ncM}). In that case, different time-dependent solutions would correspond to distinct superselection sectors on the cigar geometry. Another option is to take $\mathcal{M}_k$ to be a torus $\mathbb{T}^{n}$ with $n>22$. This leads to a supercritical string theory, which adds a positive cosmological constant to the potential. This leads to a timelike linear dilaton with slope of order one (in the $1/k$ expansion). It would be interesting to generalize our analysis to this case.

A third possibility is to take $\mathcal{M}_k=\mathbb{T}^{22}$, which results in a slightly subcritical theory. Here, a negative cosmological constant of order $1/k\alpha'$ is added to the Lagrangian. In this case there is no static solution, but one can construct time-dependent solutions. The Hamiltonian constraint now allows time-symmetric solutions, with $\dot\Phi(0)=\dot{\tilde\phi}(0)=0$ and $\tilde\phi(0)> 0$. Due to the positive gradient of the potential $V(\tilde\phi)$, this solution has the asymptotic behavior $\tilde\phi(t)\to-\infty$ as $t\to \pm \infty$, resembling the cosmology shown in figure~\ref{tphip0m01} and thus producing a big bang/big crunch singularity.

One can also generalize the analysis presented in this paper by relaxing the constraint $\chi=-\sqrt2\phi$, which further breaks the diagonal $SU(2)$ symmetry (see (\ref{Lint}) and (\ref{aaa})). Similarly, the radius deformation (\ref{fff}) can be generalized to
\begin{equation}
   \sum_{a=1}^3 \gamma_a(t)J^a  \sum_{\bar b=1}^3\bar J^{\bar b} V_1^{a\bar b}\ .
\end{equation}
This form indicates that each independent deformation $\gamma_a$ breaks the $SO(4)=SU(2)_L\times SU(2)_R$ symmetry to  $U(1)_L\times SU(2)_R$. In  spacetime, they correspond to anisotropic deformations of the three-sphere metric. See, e.g., \cite{Bakas:2006bz,Kawaguchi:2010jg,Kawaguchi:2011mz,Schubring:2022rla} for related discussions.

Finally, one can extend the background \eqref{rts3} by replacing $\mathbb{R}_t$ with a $d$ dimensional Minkowski spacetime $\mathbb{R}^{d-1,1}$. While generic solutions depend on all $d$ spacetime coordinates $(x_0,x_1,x_2,\cdots, x_{d-1})$, a particularly symmetric class preserves the $SO(d-1,1)$ isometry. To construct such solutions, we foliate spacetime into hyperbolic slices and consider configurations that depend only on the foliation (radial) coordinate. Outside the light cone, these solutions are identical to the spherically symmetric solutions on $\mathbb{R}^d$ studied in \cite{Chu:2025kzl}, which can be seen by Wick-rotating the time coordinate within the hyperbolic slices to obtain $(d-1)$-spheres. Inside the light cone, the corresponding solutions describe an expanding bubble of condensate, as noted in \cite{Chen:2021dsw}.

\section*{Acknowledgements}
We thank Ofer Aharony, Micha Berkooz, Sunny Itzhaki, Massimo Porrati and Fengjun Xu for helpful discussions. JC thanks New York University, Simons Center for Geometry \& Physics, 
Beijing Institute of Mathematical Sciences \& Applications, and Institute of Theoretical Physics, Chinese Academy of Sciences for hospitality during the conclusion of this work. DK thanks Tel Aviv University and the Weizmann Institute for hospitality. This work was supported in part by DOE grant DE-SC0009924. The work of JC was further supported in part by DOE grant 5-29073. 

\appendix
\section{Climbing up an exponential potential with anti-friction}\label{exp}
In this appendix, we generalize the cases studied in the main text by systematically analyzing anti-damped motion in the space of a field $f$, subject to an asymptotically exponential potential of the form
\begin{equation}
\label{Wf}
    W(f)\sim e^{\alpha f},\quad f\to+\infty\ .
\end{equation}
In particular, we derive a critical value for the exponent $\alpha$ that determines whether the field $f$ can escape to $+\infty$.

The relevant e.o.m. is
\begin{equation}
\label{eomW}
    \ddot f=2\dot\Phi\dot f-\frac{1}{6}W'(f)\ ,
\end{equation}
with an anti-friction coefficient given by the Hamiltonian constraint
\begin{equation}
\label{dPhiW}
    \dot\Phi=\frac{1}{2}\sqrt{3\dot f^2+W(f)}\ .
\end{equation}
Note that (\ref{tphieom}) and (\ref{H0}) are recovered with the identifications $f=\tilde\phi$ and $W=\frac{2}{k}V$. In that case, the potential form in (\ref{LKVtphi}) yields an exponent $\alpha=6$. Similarly, setting $f=-\sigma$ and $W=U$ reduces (\ref{eomW}) and (\ref{dPhiW}) to the first equation in (\ref{sigeom}) and to (\ref{H0sig}), respectively. From (\ref{Usig}), the corresponding exponent is also $\alpha=6$.

In the large $f$ regime, substituting the asymptotic form (\ref{Wf}) and the Hamiltonian constraint (\ref{dPhiW}) into (\ref{eomW}) yields
\begin{equation}
\label{eomW2}
    \dot f\frac{d\dot f}{df}=\sqrt{3\dot f^2+e^{\alpha f}}\dot f-\frac{\alpha}{6}e^{\alpha f}\ .
\end{equation}
Defining $u\equiv \dot fe^{-\alpha f/2}$, we further simplify (\ref{eomW2}) to
\begin{equation}
    u\frac{du}{df}+\frac{\alpha}{2}u^2=\sqrt{3u^2+1}u-\frac{\alpha}{6}\ ,
\end{equation}
or, equivalently,
\begin{equation}
\label{dfdu}
    \frac{df}{du}=\frac{u}{\sqrt{3u^2+1}u-\frac{\alpha}{2}u^2-\frac{\alpha}{6}}\ .
\end{equation}

For a field $f$ climbing the potential ($\dot f>0$), we have $u>0$ by definition. Then, if $\alpha\ge 2\sqrt3$, the denominator on the r.h.s. of (\ref{dfdu}) is negative for all $u>0$. Consequently, $f$ reaches its maximum at $u=0$. Since the r.h.s. of (\ref{dfdu}) vanishes at $u=0$, this maximal value of $f$ is finite. Hence, $f$ cannot reach $+\infty$. This explains why the fields $\tilde\phi$ and $\sigma$, associated with $\alpha=6$, cannot develop singularities at $\tilde\phi=+\infty$ and $\sigma=-\infty$.

On the other hand, for $0<\alpha<2\sqrt3$, the r.h.s. of (\ref{dfdu}) is negative when $0<u<u_0$, but positive when $u>u_0$, with a simple pole at
\begin{equation}
    u_0=\frac{\alpha}{\sqrt{36-3\alpha^2}}\ .
\end{equation}
In particular, at large $u$ the integration yields
\begin{equation}
    f\sim \frac{2}{2\sqrt3-\alpha}\ln u\ .
\end{equation}
This enables $f$ to escape to $+\infty$ as $u\to +\infty$.

\vskip 2cm

\bibliographystyle{JHEP}
\bibliography{HP2}

\end{document}